\documentstyle{article}[12pt]
\topmargin 10pt \textheight 225mm \textwidth 160mm \vspace{1.8cm}
\begin{document}

\baselineskip=24pt

\title{Spread theory of the special theory of relativity}
\author{X. Y. Wu$^{1}$, \thanks{The mailing address:
yxg3081@etang.com}, Xin-guo
Yin$^{1}$, Yi-Qing Guo$^{1}$ and Xin-Song Wang$^{2}$ \\
{\footnotesize 1. Department of Physics, The Nature
University, 010000, China} \\
{\footnotesize 2. Department of Mathematics, The Nature
University, 010000, China}}

\date{}
\maketitle

\renewcommand{\thesection}{Sec. \Roman{section}} \topmargin 10pt
\renewcommand{\thesubsection}{ \arabic{subsection}} \topmargin 10pt
{\vskip 5mm
\begin {minipage}{140mm}
\centerline {\bf Abstract}
\vskip 8pt
\par
\indent

The transformation of space-time $x_{\mu}$ and
${x_{\mu}}^{\prime}$ in the two inertial reference frames $\sum$
and ${\sum}^{\prime}$ in which their relative velocity is less
than light speed, and the relation of a particle mass $m$ with its
movement velocity $v$ and so on are expatiated by Einstein's
special theory of relativity. In this paper, we set forth a new
transformation of space-time $x_{\mu}$ and ${x_{\mu}}^{\prime}$ in
two inertial reference frames in which their relative velocity is
equal to or more than light speed, the new relation of a particle
mass $m$ and energy $E$ with its velocity $v$ $(v \geq c)$, the
new mass-energy equation and the new dynamics equation of a
particle.

\end {minipage}

\vspace*{2cm} {\bf PACS number(s): 03. 30. +p}
\newpage
\section * {1. Introduction}
Einstein's special relativity modify all the laws of physics,
where necessary, so as to make them a metaprinciple : it puts
constraints on all the lows of physics. The modifications
suggested by the theory, though highly significant in many modern
applications, have negligible effect in most classical problems,
which is of course why they were not discovered earlier. However,
they were not exactly needed empirically in 1905 either. This is a
beautiful example of the power of pure thought to leap ahead of
the empirical frontier-a feature of all good physical theories,
though rarely on such a heroic scale. It has led, among other
things, to a new theory of space and time, and in particular to
the relativity of simultaneity and the existence of a maximum
speed for all particles and signals, to a new mechanics in which
mass increases with speed, to the formula $E=mc^2$ $[1-5]$. The
special theory of relativity is based on two postulates formulated
by Einstein $[5]$: \\
1. All laws of nature are the same in all inertial reference
frames. In other words, we can say that the equations expressing
the laws of nature are invariant with respect to transformations
of coordinates and time from one inertial reference frame to
another.\\
2. light always propagates in a vacuum at a definite constant
speed $c$ not depending on the state of motion of the emitting body.\\
Einstein's two assumption are all right in all inertial reference
frames in which their relative velocity $v$ is less than light
speed, because all the results are proved by the experiment.
However, when two inertial reference frames relative velocity $v$
is $v=c$ or $v>c$, the two assumption in the above don't hold. For
example, when a beam of photons move in the same direction, all
photons relative velocity $v$ is zero and isn't light velocity
$c$. So, the assumption about the invariant of the light speed is
incorrect in this situation, and the result that light velocity
$c$ is a maximum speed is also incorrect, because the relative
velocity of two beams of lights moving along opposite direction
exceeds light speed $c$. So, when two inertial reference frames
relative velocity $v$ is equal to or more than light speed
Einstein's two postulations and some results must be modified.\\

\section * {2. The space-time relation at $v=c$}

Consider two inertial reference frames $\sum$ and $\sum^{\prime}$
in which their relative velocity is $v$ ($v=c$). We shall select
the coordinate axes of these frames so that the axes $x$ and
$x^{\prime}$ are directed along the velocity ${\bf v}$ of the
frame $\sum^{\prime}$, and the axes $y$ and $z$ are parallel to
the axes $y^{\prime}$ and $z^{\prime}$. Let us assume that at the
time $t=0$ a electromagnetic wave leaves the origin of $\sum$ and
$\sum^{\prime}$, which propagates with a speed $c$ in all
directions in inertial reference frames $\sum$, but it propagates
with different velocity in different direction in inertial
reference frames $\sum^{\prime}$. So, the electromagnetic wave is
a spherical wave in $\sum$, and it is a ellipsoid wave in
$\sum^{\prime}$. That is,
\begin{equation}
c^2t^2-x^2-y^2-z^2=0
\end{equation}
\begin{equation}
\frac{(x^{\prime}+\frac{ac{t^{\prime}}}{2})^2}{(\frac{ac{t^{\prime}}}{2})^2}+
\frac{(y^{\prime})^2}{(bct^{\prime})^2}+\frac{(z^{\prime})^2}{(bct^{\prime})^2}=1
\end{equation}
where $ac$ is the light velocity along with $x^{\prime}$ axes'
negative direction, and the light velocity is zero in $x^{\prime}$
axes positive direction, and $bc$ is the light velocity along with
$y^{\prime}$ and $z^{\prime}$ axes in the $\sum^{\prime}$
reference frames. \\
Obviously, the events coordinates satisfy
\begin{equation}
x=\gamma^{\prime}(x^{\prime}+ct^{\prime})
\end{equation}
\begin{equation}
x^{\prime}=\gamma(x-ct)
\end{equation}
\begin{equation}
y=y^{\prime}, z=z^{\prime}
\end{equation}
From Eqs. (3)-(4), we have
\begin{equation}
t^{\prime}=\gamma[t-\frac{x}{c}(1-\frac{1}{\gamma\gamma^{\prime}})]
\end{equation}

and from Eqs. (1)-(6), we obtain
\begin{eqnarray}
&&(4b^2\gamma^2-4ab^2\gamma^2+4ab^2\gamma^2\frac{1}{\gamma
\gamma^{\prime}})x^2+(8ab^2c\gamma^2-8b^2c\gamma^2-4ab^2c\gamma
\frac{1}{\gamma^{\prime}})xt+(4b^2c^2\gamma^2-
4ab^2c^2\gamma^2)t^2 \nonumber \\
&&=a^2x^2-a^2c^2t^2
\end{eqnarray}
from Eq. (7), we obtain
\begin{equation}
a^2=4b^2\gamma^2-4ab^2\gamma^2+4ab^2\frac{\gamma}{\gamma^{\prime}}
\end{equation}
\begin{equation}
8ab^2c\gamma^2-8b^2c\gamma^2-4ab^2c\frac{\gamma}{\gamma^{\prime}}=0
\end{equation}
\begin{equation}
4b^2c^2\gamma^2-4ab^2c^2\gamma^2=-a^2c^2
\end{equation}
from Eqs. (8)-(10), we have
\begin{equation}
\frac{\gamma}{\gamma^{\prime}}=\frac{a}{2b^2}
\end{equation}

If we let
\begin{equation}
\gamma=\gamma^{\prime}, b=1
\end{equation}
then
\begin{equation}
a=2
\end{equation}
By substituting these values into Eqs. (3)-(6), we obtain,
finally, the following transformation equations of the events
coordinates and velocity
\begin{equation}
x^{\prime}=x-ct
\end{equation}
\begin{equation}
x=x^{\prime}+ct
\end{equation}
\begin{equation}
y^{\prime}=y, z^{\prime}=z
\end{equation}
\begin{equation}
t^{\prime}=t
\end{equation}
\begin{equation}
v_{x}^{\prime}=v_{x}-c
\end{equation}
\begin{equation}
v_{y}^{\prime}=v_{y}, v_{z}^{\prime}=v_z
\end{equation}
for the general a and b, we have
\begin{eqnarray}
\gamma=\gamma^{\prime}=\sqrt{\frac{a}{2(a-b)}}
\end{eqnarray}
\begin{equation}
x^{\prime}=\sqrt{\frac{a}{2(a-b)}}(x-ct)
\end{equation}
\begin{equation}
x=\sqrt{\frac{a}{2(a-b)}}(x^{\prime}+ct^{\prime})
\end{equation}
\begin{equation}
t^{\prime}=\sqrt{\frac{a}{2(a-b)}}(t-\frac{x}{c}(\frac{2b}{a}-1))
\end{equation}
\begin{equation}
v_{x}^{\prime}=\frac{v_{x}-c}{1-\frac{1}{c}(\frac{2b}{a}-1)v_{x}}
\end{equation}
\begin{equation}
v_{y}^{\prime}=\frac{\sqrt{\frac{2(a-b)}{a}}}{1-\frac{1}{c}(\frac{2b}{a}-1)v_x}
\end{equation}
\begin{equation}
v_{z}^{\prime}=\frac{\sqrt{\frac{2(a-b)}{a}}}{1-\frac{1}{c}(\frac{2b}{a}-1)v_x}
\end{equation}
from Eq. (18), when
\begin{equation}
v_{x}=-c
\end{equation}
then
\begin{equation}
v_{x}^{\prime}=-2c
\end{equation}
from Eq. (28), we find the phenomenon exceeding light velocity can
happen in the reference frames of photons movement. This is
different from Einstein's theory of special relativity, which
light speed $c$ is the maximum velocity in the nature.
\section * {3. The space-time relation at $v>c$}

We consider two inertial reference frames $\sum$ and
$\sum^{\prime}$ in which their relative motion is the same as
section 2, and the only different is the relative velocity $v>c$.
We can obtain the following equations
\begin{equation}
x^2+y^2+z^2-c^2t^2=0
\end{equation}
\begin{equation}
\frac{(x^{\prime}+ \frac{act^{\prime}+fct^{\prime}}{2})^2}
{(\frac{act^{\prime}-fct^{\prime}}{2})^2}+
\frac{(y^{\prime})^2}{(bct^{\prime})^2}+ \frac{(z^{\prime})^2}
{(bct^{\prime})^2}=1
\end{equation}
where $ac$ is the amplitude of light velocity along the
$x^{\prime}$ axes negative direction in the $\sum^{\prime}$
reference frames which corresponds to the light velocity $-c$
along the $x$ axes negative direction in the $\sum$ reference
frames, and $fc$ is the amplitude of light velocity along the
$x^{\prime}$ axes negative direction in the $\sum^{\prime}$
reference frames which corresponds to the light velocity $c$ along
the $x$ axes positive direction in the $\sum$ reference frames,
and $bc$ is the light velocity along with $y^{\prime}$ and
$z^{\prime}$ axes in $\sum^{\prime}$ reference frames.

The events coordinates in $\sum$ and $\sum^{\prime}$ reference
frames satisfy following equations
\begin{equation}
x=\gamma^{\prime}(x^{\prime}+et^{\prime})
\end{equation}
\begin{equation}
x^{\prime}=\gamma(x-et)
\end{equation}
\begin{equation}
y^{\prime}=y, z^{\prime}=z
\end{equation}
where $e=v$ $(v>c)$, from Eqs. (31)-(32), we obtain
\begin{equation}
t^{\prime}=\gamma[t-\frac{x}{e}(1-\frac{1}{\gamma\gamma^{\prime}})]
\end{equation}
and from Eqs. (29)-(34), we have
\begin{eqnarray}
&&4b^2{\gamma}^2[(1-(a+f)c\frac{1}{e}+(a+f)c \frac{1}{e}
\frac{1}{\gamma\gamma^{\prime}}+\frac{afc^2}{e^2}+
\frac{afc^2}{e^2}\frac{1}{\gamma^2{\gamma^{\prime}}^2}-
\frac{2afc^2}{e^2}\frac{1}{\gamma \gamma^{\prime}})x^2\nonumber
\\&&
-(2e-2(a+f)c+(a+f)c \frac{1}{\gamma \gamma^{\prime}}+
2afc^2\frac{1}{e}-2afc^2\frac{1}{e}\frac{1}{\gamma
\gamma^{\prime}})xt \nonumber \\&& +(e^2-(a+f)ce+afc^2)t^2]
\nonumber \\&& =(a-f)^2x^2-(a-f)^2c^2t^2
\end{eqnarray}
from the Eq. (35), we can obtain
\begin{equation}
4b^2\gamma^2[1-(a+f)c\frac{1}{e}+(a+f)c\frac{1}{e}\frac{1}{\gamma
\gamma^{\prime}}+ \frac{afc^2}{e^2}+
\frac{afc^2}{e^2}\frac{1}{{\gamma}^2
{\gamma^{\prime}}^2}-\frac{2afc^2}{e^2}\frac{1}{\gamma
{\gamma}^{\prime}}]=(a-f)^2
\end{equation}
\begin{equation}
2e-2(a+f)c+(a+f)c\frac{1}{\gamma
\gamma^{\prime}}+2afc^2\frac{1}{e}-2afc^2\frac{1}{e}\frac{1}{\gamma
\gamma^{\prime}}=0
\end{equation}
\begin{equation}
4b^2{\gamma}^2[e^2-(a+f)ce+afc^2]=-(a-f)^2c^2
\end{equation}
from the Eqs. (36)-(38), we find
\begin{equation}
{\gamma}^2=\frac{(a-f)^2c^2}{4b^2[(a+f)ce-e^2-afc^2]}
\end{equation}
\begin{equation}
{\gamma^{\prime}}^2=\frac{4b^2afc^2(e^2-(a+f)ce+afc^2)}
{(a-f)^2(e^4-ace^3-fce^3+afc^2e^2-e^2c^2+fc^3e+ac^3e-afc^4)}
\end{equation}
from Eqs. (31)-(34), we obtain
\begin{equation}
{v_x}^{\prime}=\frac{v_x-e}{1-\frac{1}{e}v_x+\frac{1}{e}\frac{1}
{\gamma \gamma^{\prime}}v_x}
\end{equation}
when
\begin{equation}
v_x=-c
\end{equation}
then
\begin{eqnarray}
{v_x}^{\prime}=\frac{-c-e}{1+\frac{c}{e}-
\frac{c}{e}\frac{1}{\gamma \gamma^{\prime}}}=-ac
\end{eqnarray}
and when
\begin{equation}
v_x=c
\end{equation}
then
\begin{eqnarray}
{v_x}^{\prime}=\frac{c-e}{1-\frac{c}{e}+\frac{c}{e}\frac{1}{\gamma
\gamma^{\prime}}}=-fc
\end{eqnarray}
from Eqs. 39-45, we have
\begin{equation}
e^2{\gamma^{\prime}}^2=b^2{\gamma}^2e^2{\gamma^{\prime}}^2-
b^2\gamma^2{\gamma^{\prime}}^2c^2+2c^2\gamma\gamma^{\prime}-c^2
\end{equation}
for $e>c$, the Eq. (46) is identical equation. So, we have
\begin{equation}
b^2\gamma^2=1
\end{equation}
\begin{equation}
-b^2\gamma^2{\gamma^{\prime}}^2c^2-c^2+2c^2\gamma\gamma^{\prime}=0
\end{equation}
from the Eqs. (47)-(48), we find
\begin{equation}
\gamma=\gamma^{\prime}=1, b=1
\end{equation}
By substituting these values into Eqs. (31)-(34), we can obtain
the transformation equations of the events coordinates and
velocity in the $\Sigma$ and $\Sigma^{\prime}$ reference frames.
\begin{equation}
x=(x^{\prime}+et^{\prime})
\end{equation}
\begin{equation}
x^{\prime}=(x-et)
\end{equation}
\begin{equation}
y^{\prime}=y, z^{\prime}=z
\end{equation}
\begin{equation}
t=t^{\prime}
\end{equation}
\begin{equation}
{v_x}^{\prime}=v_x-e
\end{equation}
\begin{equation}
{v_y}^{\prime}=v_y, {v_z}^{\prime}=v_z
\end{equation}

Now, we can obtain the following result: when the relative
velocity of two inertial reference frames $\sum$ and
$\sum^{\prime}$ is $v=c$ and $v>c$ they have the same
transformation relation about space-time, and they are different
from the Lorentz transformation in the special theory of
relativity.

\section * {4. Mass and energy of a particle}

In special theory of relativity, Einstein found the relation of a
particle mass and energy with its motion velocity $v$, which is
less then light speed $c$. In this section, when a particle motion
velocity $v$ is large than light speed $c$ or equal to $c$, the
relation of its mass and energy with its velocity $v$ will be
established. We consider the collision of two particles which they
are identical particle. The $\sum$ is laboratory frame, and the
$\sum^{\prime}$ is the mass center frame, and two inertial
reference frame relative velocity is $c$. Before colliding, two
particle' velocity is $d$ $(d>c)$ and $v_2$ in the $\sum$ frame,
and it is $-v^{\prime}$ and $v^{\prime}$ in the $\sum^{\prime}$
frame. After colliding, all particle' velocity is $c$, and their
velocity are zero in
$\sum^{\prime}$ frame. \\
from the law of momentum conservation, we have
\begin{equation}
m_1d+m_2v_2=(m_1+m_2)c
\end{equation}
from Eq. (18), we obtain
\begin{equation}
d=v^{\prime}+c
\end{equation}
\begin{equation}
v_2=-v^{\prime}+c
\end{equation}
By substituting Eqs. (57)-(58) into Eq. (56), we have
\begin{equation}
m_1(d)=m_2(v_2)
\end{equation}
\begin{equation}
d=2c-v_2
\end{equation}
from Eqs. (59)-(60), we obtain
\begin{eqnarray}
m_1(d)=m_2(v_2)= \frac{m_0}{\sqrt{1-\frac{v_2^2}{c^2}}}
=\frac{m_0}{\sqrt{1-\frac{(2c-d)^2}{c^2}}}
\end{eqnarray}
So, when a particle has movement velocity $d$ $(d>c)$, its mass
is:
\begin{equation}
m=\frac{m_0}{\sqrt{1-\frac{(2c-d)^2}{c^2}}}
\end{equation}
where $m_0$ is its rest mass. In the following, we calculate a
particle mass when it moves at light speed $c$. It is similar to
calculating $m(d)$ $(d>c)$ the above. Before colliding, two
particle's velocity are $d$ $(d>c)$ and $-c$ in the $\sum$ frame,
and they are $-f^{\prime}$ and $f^{\prime}$ in the $\sum^{\prime}$
frame respectively. After colliding, all particle' velocity are
$c$, and their velocity are zero in $\sum^{\prime}$ frame. \\
In according to the law of momentum conversation, we have
\begin{equation}
m_1d-m_2c=(m_1+m_2)c
\end{equation}
from Eq. (18), we obtain
\begin{equation}
d=c+f
\end{equation}
\begin{equation}
-c=c-f
\end{equation}
from Eqs. (63)-(65), we have
\begin{equation}
m_1(d)=m_2(c)
\end{equation}
\begin{equation}
d=3c
\end{equation}
So, we have
\begin{equation}
m_2(c)=m_1(d)=\frac{m_0}{\sqrt{1-\frac{(2c-d)^2}{c^2}}}=
\frac{m_0}{\sqrt{1-\frac{c^2}{c^2}}}=\infty
\end{equation}
from the special theory of relativity, the relation of a particle
mass with its velocity $v$ $(v<c)$ is:
\begin{equation}
m(v)=\frac{m_0}{\sqrt{1-\frac{v^2}{c^2}}}
\end{equation}
from the Eqs. (62), (68), and (69), we get the following results:
when a particle's velocity $v$ is less then light speed $c$, its
mass $m$ is increased as its speed $v$ increases. when a
particle's velocity is $c$, its mass is infinity. when a
particle's velocity $v$ is larger then light velocity $c$, its
mass $m$ is decreased as its speed $v$ increases. when a
particle's velocity $v$ is $v>3c$, its mass is imaginary number.

In the following, Let's consider the relation of a particle energy
with the amplitude of its velocity $\overrightarrow{V}$ $(V>c)$,
and mass-energy equation, and its dynamics equation.\\
a particle's momentum $\overrightarrow{P}$ and exerted force
$\overrightarrow{F}$ are defined as:
\begin{equation}
\overrightarrow{P}=\frac{m_0{\overrightarrow{V}}}
{\sqrt{1-\frac{(2c-V)^2}{c^2}}}
\end{equation}
\begin{equation}
\overrightarrow{F}=\frac{d}{dt}\overrightarrow{P}
\end{equation}
and
\begin{equation}
\frac{d}{dt}w=\overrightarrow{F}\cdot\overrightarrow{V}
\end{equation}
where $w$ is a particle energy. from the Eq. (70)-(72), we have
\begin{equation}
w=\frac{m_0c^2}{\sqrt{1-\frac{(2c-V)^2}{c^2}}}\frac{(3c-2V)V}{(2c-V)c}
\end{equation}
and mass-energy equation
\begin{equation}
P^2(\frac{(3c-2V)c}{2c-V})^2-w^2=0
\end{equation}
and a particle's dynamics equation
\begin{equation}
\overrightarrow{F}=\frac{m_0}{\sqrt{1-\frac{(2c-V)^2}{c^2}}}
\frac{d}{dt}\overrightarrow{V}-\frac{m_0\overrightarrow{V}}
{(\sqrt{1-\frac{(2c-V)^2}{c^2}})^3}\frac{2c-V}{c^2V}\overrightarrow{V}
\cdot\frac{d}{dt}\overrightarrow{V}
\end{equation}
\section * {5. conclusion}
To summarize, we have discussed the time-space transformation in
two inertial reference frames which their relative velocity is
equal to or more than light speed and researched the relation of a
particle's mass and energy with its motion velocity and its
dynamics equation. It is different from the special theory of
relativity, which based on Einstein's principle of relativity and
the principle of constancy of the speed of light. When two inertia
reference frames' relative velocity and a particle's velocity is
less than light velocity they are described by the special theory
of relativity. However, when two inertial reference frames'
relative velocity and a particle's velocity is equal to or larger
than light velocity it must be described by the spread theory of
the special theory of relativity.


\begin{thebibliography}{99}


\bibitem{s1}
W. G. Dixon, Special Relativity, Cambridge University, 1978, pp.
1-76.

\bibitem{s2}
Wolfgang Rindler, Essential Relativity, the United States of
America, 1977, pp. 1-125.

\bibitem{s3}
Wolfgang Rindler, Introduction to Special Relativity, Oxford
University, 1982, pp. 1-112.

\bibitem{s4}
Robert Resnick, Introduction to Special Relativity, the United
States of America, 1968, pp. 1-148.

\bibitem{s5}
I. V. Savelyev, Fundamentals of Theoretical Physics, Frist
published 1982 and revised from the 1975 Russian edition, pp.
125-156.

\end{thebibliography}
\end{document}